\def\deg{\mbox{$^{\circ}$}}

\newcommand\JMR{ {J.~Magn.~Reson.} }
\newcommand\JMRA{ {J.~Magn.~Reson.~Series A} }

\newcommand\MRM{ {Magn.~Reson.~Med.} }
\newcommand\vol[1]{{#1}}

\newcommand\bm[1]{\mbox{\boldmath $#1$}}
\newcommand\Eq[1]{Eq.$\:$[\,\ref{#1}\,]}
\newcommand\Fig[1]{Fig.$\:$\ref{#1}}
\newcommand\Figure[1]{Figure$\:$\ref{#1}}

\newcommand\Cthteen{$^{13}$C}
\newcommand\musec{$\mu$s}
\newcommand\PPinv{$180^\circ_{PP}$}
\newcommand\URinv{$180^\circ_{UR}$}
\newcommand\PPexc{$90^\circ_{PP}$}
\newcommand\URexc{$90^\circ_{UR}$}
\newcommand\BURBOPinv{$180^\circ_{BURBOP}$}

\newcommand\T[1]{\mbox{$T_{\rm #1}$}}

\documentclass[preprint,12pt]{elsarticle}

\usepackage{amssymb}
\usepackage{graphicx}
\usepackage{lineno}

\begin{document}

\begin{frontmatter}



\title{Broadband 180\deg\ universal rotation pulses \\ 
       for NMR spectroscopy designed by optimal control }


\author[a]{Thomas E. Skinner\corref{cor1}} \ead{thomas.skinner@wright.edu}\cortext[cor1]{Corresponding author.}
\author[a]{Naum I. Gershenzon}
\author[b]{Manoj Nimbalkar}
\author[c]{Wolfgang Bermel}
\author[d,e]{Burkhard Luy}
\author[b]{Steffen J. Glaser} \ead{glaser@ch.tum.de}

\address[a]{Physics Department, Wright State University, Dayton, OH 45435, USA}
\address[b]{Department of Chemistry, Technische Universit\"at  M\"unchen, Lichtenbergstr. 4, 85747 Garching, Germany}
\address[c]{Bruker BioSpin GmbH, Silberstreifen 4, 76287 Rheinstetten, Germany}
\address[d]{Institut f\"ur Organische Chemie, Karlsruher Institut f\"ur Technologie, Fritz-Haber-Weg 6, 76131 Karlsruhe, Germany}
\address[e]{Institut f\"ur Biologische Grenzfl\"achen 2, Karlsruher Institut f\"ur Technologie, 76344 Karlsruhe, Germany}

\begin{abstract}
Broadband inversion pulses that rotate all magnetization components 180\deg\ about a given fixed axis are necessary for refocusing and mixing in high-resolution NMR spectroscopy. The relative merits of various methodologies for generating pulses suitable for broadband refocusing are considered.  The de novo design of 180\deg\ universal rotation pulses (\URinv) using optimal control can provide improved performance compared to schemes which construct refocusing pulses as composites of existing pulses.   The advantages of broadband universal rotation by optimized pulses (BURBOP) are most evident for pulse design that includes tolerance to RF inhomogeneity or miscalibration.  We present new modifications of the optimal control algorithm that incorporate symmetry principles and relax conservative limits on peak RF pulse amplitude for short time periods that pose no threat to the probe.  We apply them to generate a set of \BURBOPinv\ pulses suitable for widespread use in \Cthteen\ spectroscopy on the majority of available probes.  
\end{abstract}

\end{frontmatter}

\parindent = .2truein

\section{Introduction}

Many NMR applications require refocusing of transverse magnetization, which is easily accomplished on resonance by any good inversion pulse sandwiched between delays, ie, the standard $\Delta$--180\deg--$\Delta$ block. For broadband applications, a universal rotation (UR) pulse that rotates any orientation of the initial magnetization 180\deg\ about a given fixed axis is required to refocus all transverse magnetization components.  
A simple hard pulse functions as a UR pulse only over a limited range of resonance offsets that can not be increased significantly due to pulse power constraints.

Although a great deal of effort has been devoted to increasing the bandwidth of inversion pulses, most broadband inversion pulses  \cite{Levitt79,Shaka83a,Shaka83b,Pines83,Levitt83,Hoult85,Pines85,Shaka85,Levitt86,KF95,KF96,Garwood96,Hwang98,BIP,bblimits,rf-power-limits} execute only a point-to-point (PP) rotation for one specific initial state, magnetization $M_z \rightarrow -M_z$, and are not UR pulses.  However, two PP inversion pulses with suitable interpulse delays can be used to construct a refocusing sequence \cite{Levitt80,Conolly91}, which is effectively a $360^\circ_{UR}$ pulse.  Alternatively, a \URinv\ refocusing pulse can be constructed from three adiabatic inversion pulses \cite{Hwang97} with either pulse length or bandwidth of the adiabatic frequency sweep in the ratio 1:2:1.  More generally, we have shown that one can construct a UR pulse of any flip angle from a PP pulse of half the flip angle preceded by its time- and phase-reversed waveform \cite{URconstruction}. Thus, 
a \URinv\ pulse can be constructed from two \PPexc\ pulses.

The reliance on composites of PP pulses to construct UR pulses highlights the perceived difficulty of creating stand-alone UR pulses.  The de novo design of UR pulses for NMR spectroscopy has received comparatively little attention \cite{Levitt86, Shaka87}, so it is an open question whether the composite constructions using PP pulses achieve the best possible performance.
Yet, the demonstrated capabilities of optimal control for designing PP pulses \cite{Conolly86, Mao86, James91a, James91b, Rosenfeld96, Skinner03, Skinner04, Skinner05, Skinner06, Gershenzon07, Gershenzon08, OCAM, Skinner10, Skinner11} are equally applicable to the design of UR pulses \cite{ Grape, KyrylThesis, Cory10}.  The required modifications to the basic optimal control algorithm are fairly straightforward \cite{Grape, Rangan01, Kosloff02} and maintain the same flexibility for incorporating tolerance to variations in experimentally important parameters, such as RF homogeneity or relaxation.

In this work, we design broadband refocusing pulses by optimizing the propagator for the required UR transformation.  The resulting $180^\circ_{BURBOP}$ pulses (broadband universal rotation by optimized pulses) are compared to existing composite refocusing pulse schemes to characterize the conditions under which one design method might be preferrable to another.  In addition, we introduce new optimal control strategies tailored to take advantage of specific opportunities available in the design of UR pulses.  The culmination of these efforts is a set of low-power, high-performance broadband refocusing pulses that satisfy the power constraints of widely available probeheads and complex multipulse sequences.  

\section{Optimal control algorithm for \URinv\ pulses}

A general procedure for creating a desired unitary propagator in an arbitrary (closed) quantum system is given in \cite{Grape, Rangan01, Kosloff02, Tosh05}.  Time evolution proceeds according to a matrix exponential of the system Hamiltonian.  For two-level systems, as in many NMR applications involving a single noninteracting spin-1/2 species, this evolution is well-known to be equivalent to a rotation of the 3D vector representing the state of the system about the effective applied field \cite{Feynman}.  The relatively abstract general procedure for propagator optimization can be made considerably more transparent in this case.  

\subsection{Flavor I (basic vanilla)}
The optimal control methodology for generating PP transformations in two-level systems has been described in detail previously \cite{Conolly86, Mao86, James91a, James91b, Rosenfeld96, Skinner03, Skinner04, Skinner05, Skinner06, Gershenzon07, Gershenzon08, OCAM, Skinner10, Skinner11}, with progressive modification to enhance pulse performance by incorporating various experimental constraints.  In each iteration of the algorithm, one starts with a given initial magnetization $\bm{M}_0$, applies the RF derived for the current iteration, and compares the resulting final state $\bm{M}_f$ with a desired target state $\bm{F}$.  This comparison, quantified in terms of a cost function, allows one to efficiently calculate a gradient for improving pulse performance in the next iteration.  A simple but effective cost function $\Phi$, which we employ in the present work, is the projection of the final state onto the target state, the standard inner product $\langle\,F\,|\,M_f\,\rangle$, given in this case by the dot product $\bm{F}\cdot\bm{M}_f$.  Desired RF limits can be enforced by clipping without degrading the performance of the optimization \cite{Skinner04}.

The algorithm for generating UR pulses in the single-spin case is a straightforward modification of the PP algorithm. 
A \URinv\ pulse applied, for example, along the y-axis to magnetization $\bm{M}$ effects the transformation $(M_x, M_y, M_z) \rightarrow (-M_x, M_y, -M_z)$.  This is simply three separate PP transformations of the initial states $\bm{M}_1 = (1, 0, 0)$, $\bm{M}_2 = (0, 1, 0)$, $\bm{M}_3 = (0, 0, 1)$ to their respective target states $\bm{F}_1 = (-1, 0, 0)$, $\bm{F}_2 = (0, 1, 0)$, $\bm{F}_3 = (0, 0, -1)$.  The cost function comparing the final states $\bm{M}_{kf}$ ($k = 1, 2, 3$) at the end of an RF pulse to the target states is 
	\begin{equation}
\Phi = \bm{F}_1\cdot \bm{M}_{1f} + \bm{F}_2\cdot \bm{M}_{2f} +\bm{F}_3\cdot \bm{M}_{3f}.
\label{Cost}
	\end{equation}
The algorithm proceeds in the standard fashion using this cost function. We will refer to this as algorithm A.

This simple intuitive modification to the cost is exactly equivalent to an analogous
procedure given in \cite{Grape} for optimizing the unitary propagator, which can be seen as follows.  The rotation operator $R_F$ in 3D corresponding to the target propagator that generates a 180\deg\ rotation about the $y$-axis is given by
	\begin{equation}
R_F =
\left( \begin{array}{rcr}
-1 & 0 & 0 \\
0 & 1 & 0 \\
0 & 0 & -1 
       \end{array}
\right ) = 
\left( \begin{array}{ccc}
\vdots & \vdots & \vdots \\
\bm{F}_1 & \bm{F}_2 & \bm{F}_3 \\
\vdots & \vdots & \vdots
       \end{array}
                      \right),
	\end{equation}
ie, the $i^{\rm \,th}$ column is the corresponding PP target $\bm{F}_i$.  

The actual rotation operator at the end of a pulse of length $T_p$ is
	\begin{eqnarray}
R(T_p) & = & R(T_p) \left( \begin{array}{ccc}
                      1 & 0 & 0 \\
                      0 & 1 & 0 \\
                      0 & 0 & 1 
       \end{array}
\right ) \nonumber \\
      & = & 
R(T_p) \left( \begin{array}{ccc}
          \vdots & \vdots & \vdots \\
           \bm{M}_1 & \bm{M}_2 & \bm{M}_3 \\
           \vdots & \vdots & \vdots
                 \end{array}
         \right)  \nonumber \\
      & = & \quad\quad
\left( \begin{array}{ccc}
          \vdots & \vdots & \vdots \\
           \bm{M}_{1f} & \bm{M}_{2f} & \bm{M}_{3f} \\
           \vdots & \vdots & \vdots
                 \end{array}
         \right),
	\end{eqnarray}
with the rotation transforming each column to its associated final state for the individual PP transformations.

The cost is again given by the projection of the final state onto the target state, with the inner product 
	\begin{equation}
\Phi_R = \langle\,R_F\,|\,R(T_p)\,\rangle
     = \mbox{Tr}\,[R_F^{\cal T}\,R(T_p)],
	\end{equation}
where superscript $\cal T$ denotes the transpose, and the operator Tr returns the trace (sum of diagonal elements) of its argument.  We then have 
	\begin{equation}
\hskip -18pt
R_F^{\cal T}\, R(T_p) = 
\left( \begin{array}{lll} 
\cdots \bm{F}_1 \cdots \\
\cdots \bm{F}_2 \cdots \\
\cdots \bm{F}_3 \cdots \\
       \end{array}
\right )
\left( \begin{array}{ccc}
          \vdots & \vdots & \vdots \\
           \bm{M}_{1f} & \bm{M}_{2f} & \bm{M}_{3f} \\
           \vdots & \vdots & \vdots
                 \end{array}
         \right).
	\end{equation}
The sum over diagonal elements of this matrix product gives \Eq{Cost}.

\subsubsection{Flavor II (symmetry principle)}
\label{FlavorII}
The formalism for constructing UR pulses from PP pulses \cite{URconstruction} provides additional insight for improving \BURBOPinv\ performance.  The symmetry of the construction procedure constrains the resulting rotation axis to the plane defined by the desired axis and the $z$-axis \cite{Ngo87}. 
Details of the results which follow are provided in the Appendix.  

For a UR rotation about the $x$-axis, any nonidealities in the original PP pulse (for example, due to resonance offset) shift the resulting rotation axis only in the $xz$-plane.  Given a phase deviation of magnitude $\delta\phi$ from the desired target state and rotation error $\delta\theta$ compared to the desired rotation angle,
the angle $\alpha_{180_{UR}}$ that the rotation axis in the $xz$-plane makes with the $x$-axis is small and bounded according to the relation
	\begin{equation}
\alpha_{180_{UR}} \lesssim \sqrt{(\delta\phi_{90_{PP}})^2 + (\delta\theta_{90_{PP}})^2}
\label{alpha}
	\end{equation}
for small deviations measured in radians.  

This has the effect of reducing any phase errors in the original $90^\circ_{PP}$ pulse used for the construction.  For example, a 180\deg\ rotation about any axis in the $xz$-plane transforms $I_y$ to $-I_y$ with no phase error.  There would be no phase error in the $I_x \rightarrow I_x$ transformation, either, but the amplitude of the final $x$-magnetization would decrease depending on the angle $\alpha_{180_{UR}}$, giving a $\cos(2\alpha_{180_{UR}})$ dependence.  Similarly, the $I_z \rightarrow -I_z$ transformation would have a $\cos(\alpha_{180_{UR}})$ dependence for the magnitude of the final z-magnetization.
Phase errors are thus the result of deviations, $\delta\eta$, from the ideal 180\deg\ rotation angle, which are bounded (see Appendix) according to
	\begin{equation}
\delta\eta \lesssim 2 \sqrt{(\delta\phi_{90_{PP}})^2 + (\delta\theta_{90_{PP}})^2}
\label{d_eta}
	\end{equation}
Larger amplitude errors result from both the displacement of the rotation axis from the $x$-axis and deviations from the ideal 180\deg\ rotation angle.  Phase deviations in the \PPexc\ pulse are reduced in the resulting \URinv\ pulse according to the relation
	\begin{equation}
\delta\phi_{180_{UR}} \lesssim (\delta\phi_{90_{PP}})^2 + (\delta\theta_{90_{PP}})^2.
\label{PhaseImprovement}
	\end{equation}
The reduction is quite significant, with $\delta\theta_{90_{PP}} = 1^\circ$ and $\delta\phi_{90_{PP}}$ of 10\deg, 3\deg, and 1\deg, for example, giving bounds for $\delta\phi_{180_{UR}}$ of 1.76\deg, 0.17\deg, and 0.03\deg, respectively.

For applications requiring high phase fidelity which can afford modest loss of signal intensity, we therefore incorporate the symmetry principle of the construction procedure into the optimal control algorithm.  For RF pulse components $u_x$ and $u_y$ digitized in $N$ time steps, the first half of the pulse is determined using the basic algorithm A.  The second half of the pulse is then constructed using the time-and phase-reversed components of the first half.  Phase zero for $u_x$ leaves it unaffected, while $u_y$ is inverted to give
	\begin{eqnarray}
u_x^{i + N/2} &=& u_x^{N/2 + 1 - i} \nonumber \\
u_y^{i + N/2} &=& -u_y^{N/2 + 1 - i}
\label{AlgorithmII}
	\end{eqnarray}
for $i = 1, 2, 3,\dots,N/2$.  We refer to this algorithm incorporating the symmetry of the construction principle as A$_{S}$.

\subsubsection{Flavor III (time-dependent RF limit)}    
\label{FlavorIII}

Peak RF amplitude must remain below probe limits (e.g., available
for $^{13}$C spectroscopy), but larger RF amplitudes result in improved broadband performance.  For sufficiently short time periods, we note that probe RF limits can be higher than conservative limits that protect the probe from arcing under any conditions.  Enforcing a lower probe limit for an entire pulse duration can sacrifice performance unnecessarily.  We therefore introduce a time-dependent RF field limit to allow increased RF amplitude for short time intervals and achieve improved performance. Empirically, we find that low RF limits force algorithm A to request higher RF amplitude in the middle of the pulse for improved performance.  We therefore allow a higher RF limit for a short time during the middle of the pulse.  We refer to this algorithm as A$_T$, or, if it is also combined with the symmetry principle, as A$_{S,T}$.

\section{BURBOP compared to refocusing with PP pulses}

Performance of the new \BURBOPinv\ pulses designed using optimal control are compared with previous methods for generating refocusing pulses, starting with those that can be constructed from \PPexc\ pulses.  
We then consider the refocusing performance of two \PPinv\ pulses. In both these comparisons, RF amplitudes are limited to a relatively conservative peak value of 11 kHz (22.7 \musec\ hard 90\deg\ pulse) for widespread use in $^{13}$C spectroscopy.  Finally, we present a set of \BURBOPinv\ pulses that utilize the higher RF power limits allowed for short time intervals to increase the maximum RF amplitude from 11 kHz to 15 kHz (16.7 \musec\ hard pulse) in the middle of the pulse.  The performance of these pulses is compared with the composite adiabatic pulse scheme \cite{Hwang97} implemented as a smoothed Chirp pulse \cite{Bodenhausen93, Bermel03} in standard Bruker software . 
   
\subsection{Algorithms A and A$_S$}
\subsubsection{Two \PPexc\ pulses}

A previously published \PPexc\ pulse  with exceptional performance \cite{Skinner06} was used to construct a \URinv\ pulse according to the procedure in \cite{URconstruction}.  This 1~ms constant amplitude phase-modulated excitation pulse transforms greater than 99\% of initial z-magnetization to the x-axis over a resonance offset range of 50~kHz for RF amplitude anywhere in the range 10--20~kHz. For all but the lowest RF amplitudes, the transformation is greater than 99.5\%.  Phase deviations of the excited magnetization from the x-axis are less than 2\deg--3\deg\ over almost the entire optimization window, with minor distortions in the 6\deg--9\deg\ range at the lowest RF values.

\begin{figure*}[h]
\includegraphics[trim= 0 1.7in 0 0, clip, width= 6.25in,scale=.9]{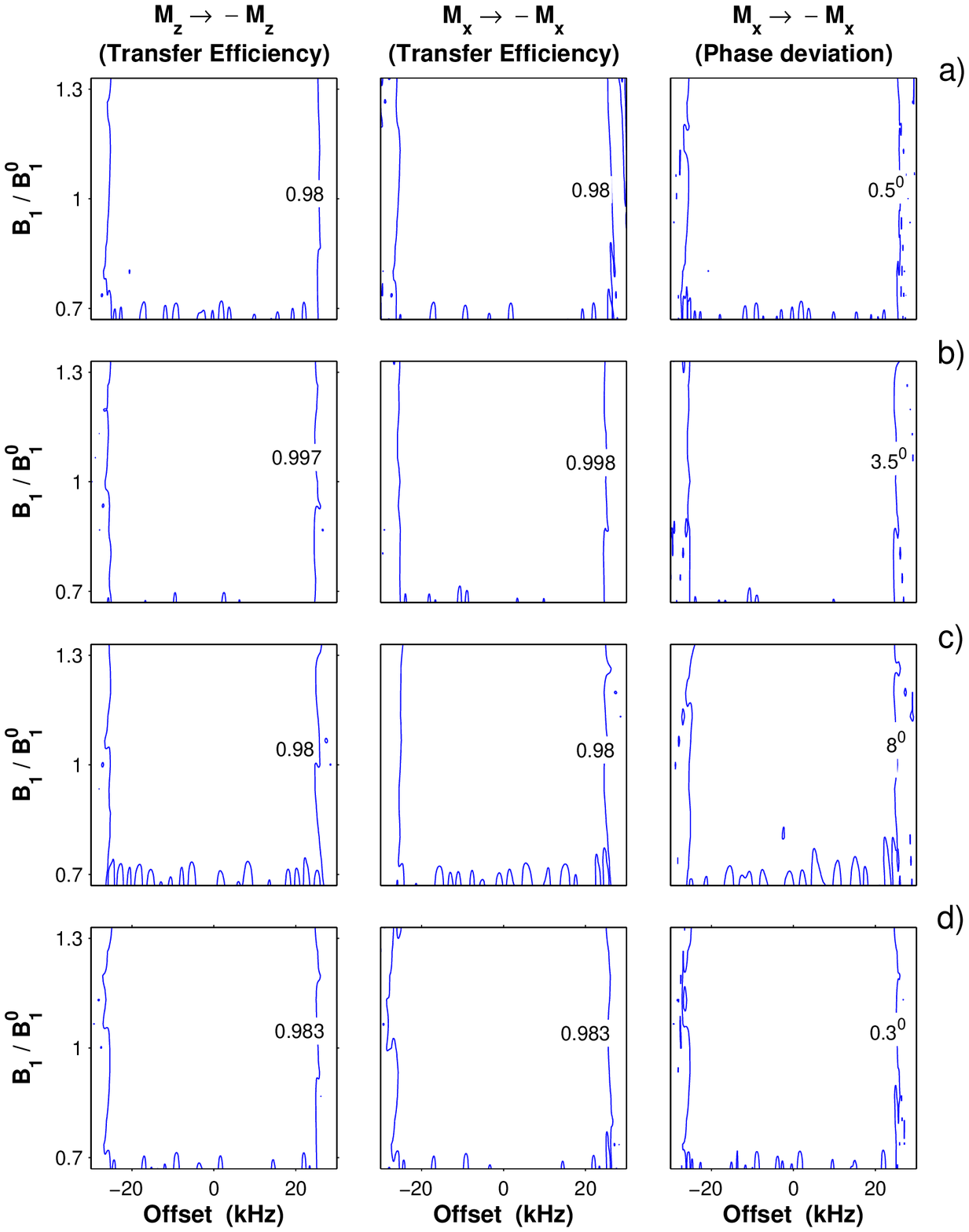}
\caption{Theoretical performance of four \URinv\ pulses for inversion of magnetization about the y-axis, designed using different algorithms.  The first two columns show the transfer efficiency for the labeled transformations relative to ideal or complete transfer.  The last column displays phase deviation in degrees relative to the labeled target state.  The nominal peak RF amplitude for all the pulses is $B_1^0 =15$~kHz, optimized to perform over a resonance offset range of 50~kHz and variation in RF homogeneity/calibration of $\pm 33\%$. All pulses are constant amplitude with the exception of pulse b), which deviates from the maximum for less than 10\%\ of the pulse.  \textbf{a)} constructed from the 1~ms \PPexc\ pulse of Ref.~\cite{Skinner06} preceded by its time- and phase-reversed waveform \cite{URconstruction}, pulse length $T_p=2$~ms. \textbf{b)} algorithm A, $T_p=2$~ms.  \textbf{c)} algorithm A, $T_p=1$~ms.  \textbf{d)} algorithm A$_S$, which incorporates the symmetry principle used in a), $T_p=2$~ms. }
\label{UR-Const}
\end{figure*}

The upper panel of \Fig{UR-Const} depicts the performance of the constructed pulse composed of two 1~ms \PPexc\ pulses.  As expected from the discussion in $\S$\ref{FlavorII}, the signal intensity of the transformed magnetization is decreased slightly compared to the original PP pulse, but the phase relative to the target axis is improved considerably.  The construction procedure rotates phase deviations relative to the target axis out of the transverse plane, reducing spectral phase errors that are the primary interest.  Increased phase dispersion is allowed in a plane orthogonal to the x-y plane, resulting in reduced projected transverse magnetization.

In comparison, the basic algorithm A, which optimizes the UR propagator, results in a pulse with performance shown in \Fig{UR-Const}b that is more consistent with the performance characteristics of the \PPexc\ pulse used for the UR construction in \Fig{UR-Const}a.  Thus, algorithm A maximizes signal amplitude, with phase fidelity secondary, while the construction procedure maximizes phase fidelity, with signal amplitude secondary.

When the optimization includes a range of RF inhomogeneity/miscalibration, algorithm A can produce shorter pulses than the construction principle for a comparable amplitude performance, as shown in \Fig{UR-Const}c, at the cost of poorer phase performance.  We find empirically that the reduction in pulse length is 50\%, 30\%, and 15\% for RF miscalibration ranges of $\pm33\%, \pm10\%$, and $\pm5\%$, respectively.  In optimizations that do not include tolerance to RF inhomogeneity, there is no reduction in pulse length compared to the construction procedure.

Finally, algorithm A$_S$ utilizes the construction procedure symmetry in optimizing the UR propagator, producing the pulse performance shown in \Fig{UR-Const}d.  Algorithm A$_S$ provides more ideal phase performance and therefore can provide an advantage over the construction procedure when tolerance to RF inhomogeneity is included in the optimization.  The choice between algorithms A and A$_S$ depends on the application, whether maximizing signal amplitude is more important than ideal phase performance.

\subsubsection{Two \PPinv\ pulses}

Two \PPinv\ pulses can be used to refocus transverse magnetization \cite{Levitt80, Conolly91} by modifying the standard spin-echo sequence. For example, a \PPexc\ excitation pulse of length $T_p$ that produces a linear phase dispersion $\Delta\omega R T_p$ ($0 \le R \le 1$) as a function of offset $\Delta\omega$ could be followed by $\tau$ - \PPinv\ - $(\tau + R T_p)$ - \PPinv.  The first $\tau$-delay already includes a phase evolution equivalent to time $R T_p$, which is included in the next delay of the standard spin-echo.  The second inversion pulse then compensates the phase errors of the first \cite{BIP}, which is unnecessary if a single \URinv\ is employed.  A shorter sequence for this example is $(\tau - R T_p)$ - \PPinv\ - $\tau$ - \PPinv.  
The resulting modified spin-echo procedures rotate magnetization 360\deg\ and are not UR inversions, but this is not an issue for refocusing.  

BIP \cite{BIP} and BIBOP \cite{bblimits, rf-power-limits} pulses are optimized to provide exceptional performance as \PPinv\ pulses, which translates to similar performance when incorporated into the above refocusing scheme.  There is little room for improving the refocusing performance of two \PPinv\ pulses. A similar conclusion was reached in the context of relatively high bandwidth selective pulses \cite{Matson09}.  On the other hand, a single \URinv\ pulse can be simpler to incorporate into complex pulse sequences with respect to adjusting the timings and synchronization among various pulses.  Conventional hard 180\deg\ pulses can be easily replaced in an existing sequence by \BURBOPinv\ pulses without further change of the pulse sequence. By contrast, incorporating two \PPinv\ pulses can require the adjustment of phases and a correspondingly detailed understanding of the pulse sequence.

We therefore find at most a modest advantage in using algorithms A or A$_S$ to generate a \URinv\ pulse compared to the overlapping spin-echo sequence using optimized \PPinv\ pulses. 
However, incorporating a time-dependent RF limit into the optimal control algorithm does
provide a distinct advantage for generating a single \URinv\ pulse compared to existing pulses, as illustrated in what follows.

\subsection{Algorithms A$_T$ and A$_{S,T}$}

Broadband refocusing bandwidths of $\sim50$~kHz, sufficient for high-field \Cthteen\ spectroscopy, are readily achieved using any of the pulse schemes discussed so far.  The peak RF power required for the pulses is well-within hard-pulse power limits for modern high resolution probes.  However, in multipulse sequences, repeated application of what might be deemed a modest power level for a single pulse can be a problem if the total energy delivered to the sample (integrated power) is too high.  There are also limits on the total energy that can safely be delivered to a given probe.
For these reasons, the most general and widespread applications impose peak power levels that are more conservative than what might be necessary for broadband refocusing using a typical probe.  We therefore incorporate a time-dependent RF limit into the optimal control algorithms to keep peak power low for most of the pulse, but allow short increases in this limit where it can have the most benefit. We utilize algorithms A$_T$ and A$_{S,T}$ (defined in $\S$\ref{FlavorIII}) to investigate the possibility of generic broadband refocusing pulses suitable for use with any standard probehead in any pulse sequence. 

\subsubsection{Three adiabatic inversion pulses}
The best broadband refocusing performance available in the standard Bruker pulse library satisfying the required conservative pulse power limits is obtained using the pulse designated Chirp80 \cite{Bermel03}.  This pulse is constructed from three adiabatic inversion pulses with pulse lengths in the ratio 1:2:1 \cite{Hwang97}.  
It utilizes for its shortest element a 500~ms smoothed chirp pulse \cite{Bodenhausen93} with 80~kHz sweep.  The first 20\%\ of the pulse rises smoothly to a maximum constant RF amplitude of 11.26~kHz according to a sine function before decreasing in the same fashion to zero during the final 20\%\ of the pulse. The final pulse is thus 2~ms long.  

\begin{figure*}[h]
\includegraphics[scale=.7]{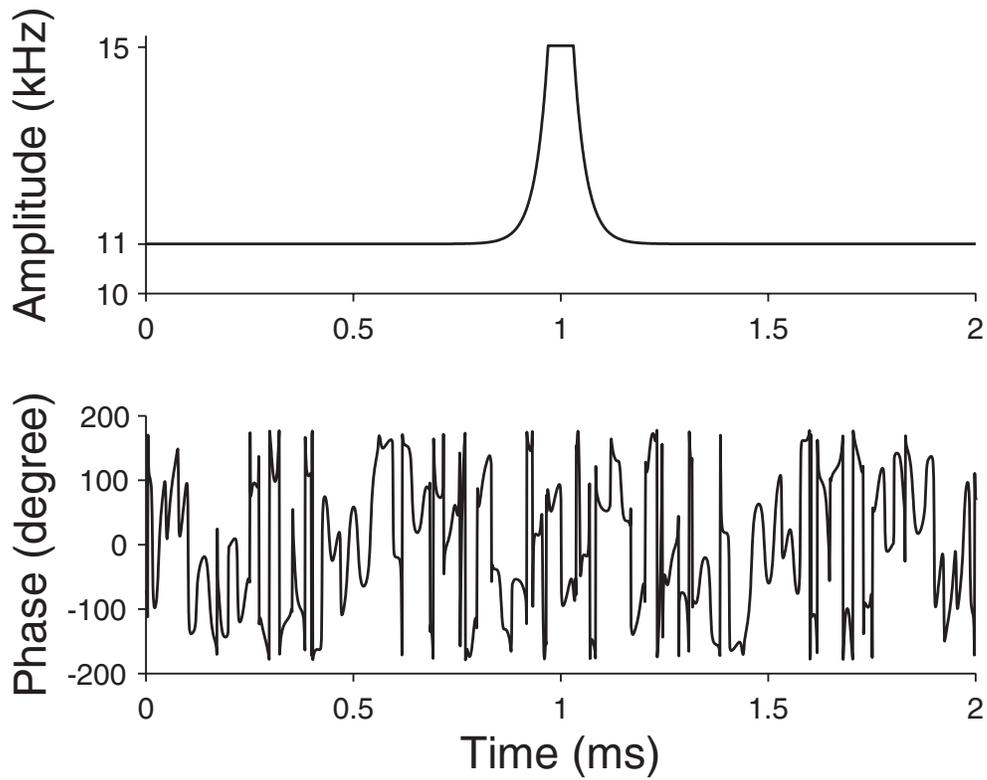}
\caption{The amplitude and phase of \URinv\ pulse 4 from Table 1 obtained using algorithm A$_{S,T}$.  A conservative limit of 11~kHz for the peak RF applied for 2~ms is relaxed to allow a safe peak of 15~kHz for 60\musec.  The pulse is amplitude-symmetric and phase-antisymmetric in time, incorporating the symmetry of the construction procedure from Ref.~\cite{URconstruction}. }
\label{Pulse4}
\end{figure*}

Maintaining this pulse length and mindful of the given conservative peak RF amplitude, we designed the set of four pulses listed in Table 1.  For most of the pulse, the nominal RF amplitude is a constant 10 or 11~kHz.  A maximum RF amplitude of 15~kHz is applied for 60 \musec\ in the middle of the pulse, as illustrated in \Fig{Pulse4}.  This short increase in pulse amplitude provides significant improvement in pulse performance compared to Chirp80.  

The amplitude profile shown in \Fig{Pulse4} is reminiscent of the hyperbolic secant pulse \cite{Hoult85}, which maintains a low amplitude for most of the pulse with a peak in the middle.  
All four pulses show excellent performance over the listed ranges in offset and RF field inhomogeneity/miscalibration.  Performance is comparable to the performance shown in \Fig{UR-Const} for higher power, constant amplitude pulses with nominal peak RF of 15~kHz.  Pulse 4 provides the most relevant comparison, since it has a similar range of tolerance to RF inhomogeneity. As expected from the earlier results for algorithms A and A$_S$, the best amplitude performance is obtained by algorithm A$_T$ and the best phase performance by algorithm A$_{S,T}$ 

\begin{figure*}[h]
\includegraphics[scale=.8]{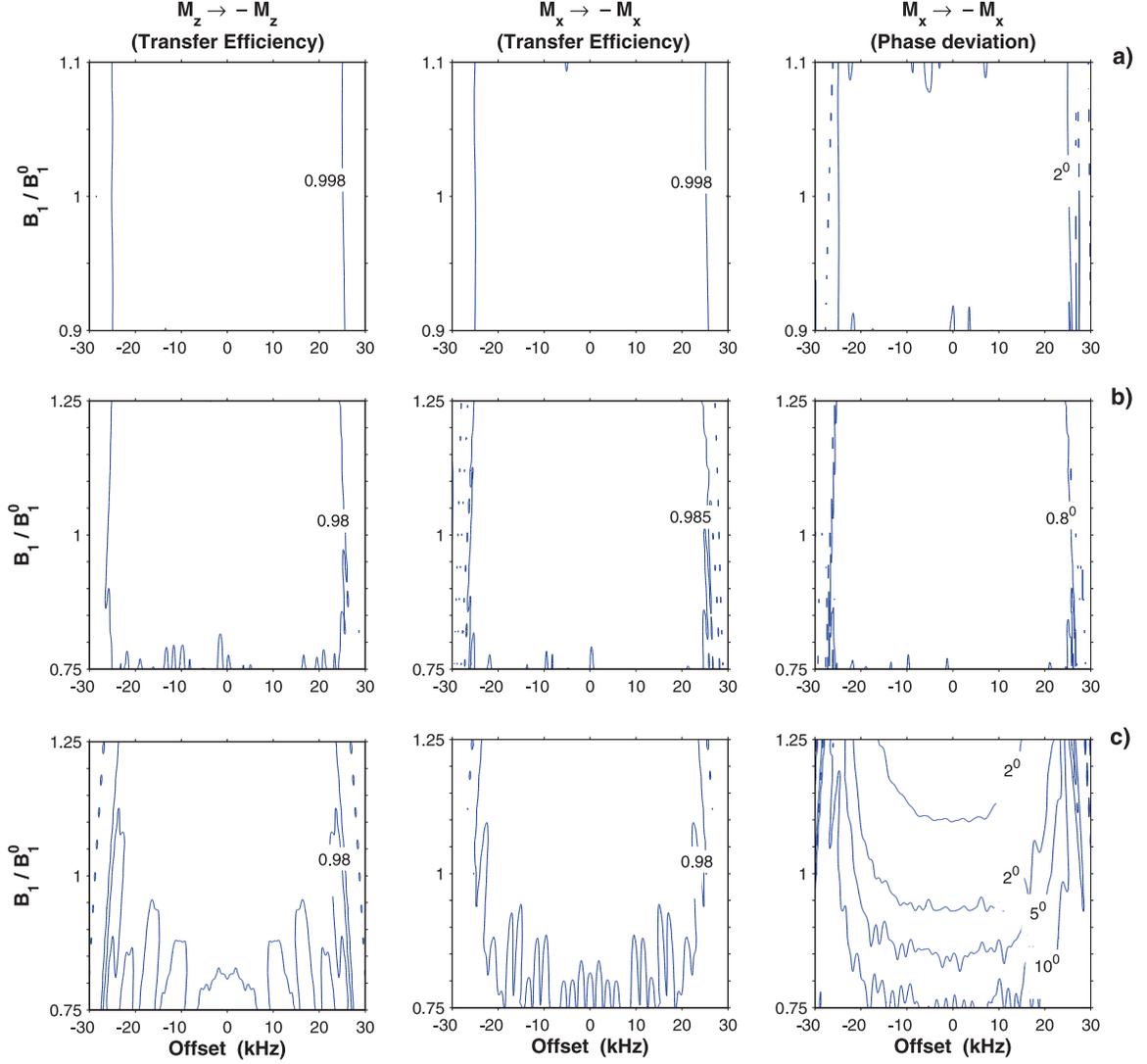}
\caption{Similar to \Fig{UR-Const}, but the maximum RF amplitude is allowed to float for a sufficiently short time period, giving the reduced-power pulse in \Fig{Pulse4}.  All three pulses are designed to operate over a resonance offset of 50~kHz and a range of variation in RF homogeneity/calibration relative to the ideal $B_1^0$, as given in Table 1.  \textbf{a)} pulse 1 of Table 1, RF tolerance $\pm10\%$.  \textbf{b)} pulse 4 of Table 1, RF tolerance $\pm25\%$.  \textbf{c)} composite adiabatic refocusing \cite{Hwang97} using pulse Chirp80 from the Bruker pulse library \cite{Bermel03}.}
\label{CompareChirp}
\end{figure*}

\begin{figure*}[h]
\includegraphics[scale=.55]{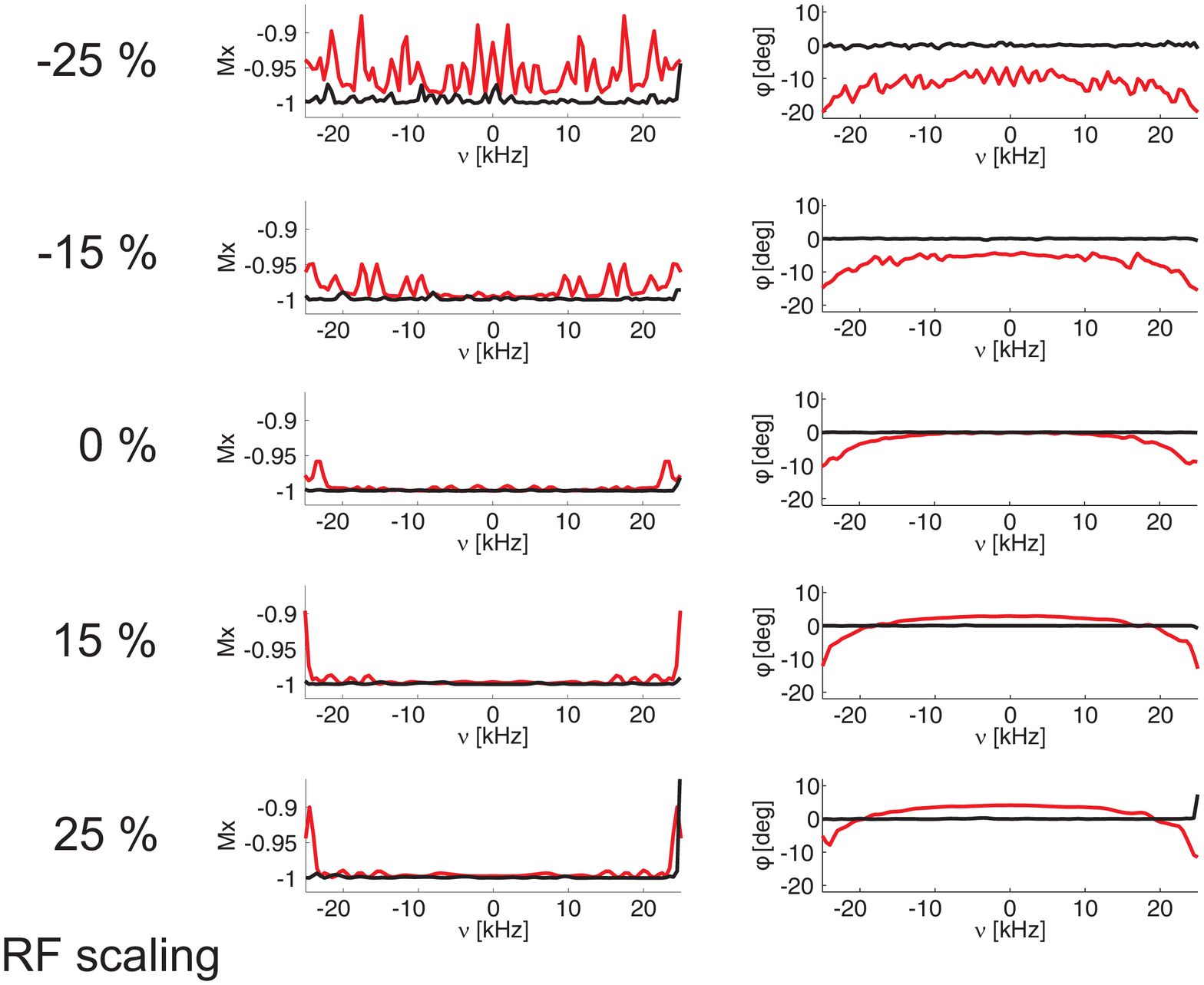}
\caption{Further quantitative detail for the $M_x\rightarrow -M_x$ transformation from \Fig{CompareChirp}b for pulse 4 of Table 1 (black) and \Fig{CompareChirp}c for Chirp80 (red), plotted for RF scalings of $\pm15\%$ and $\pm25\%$ relative to the nominal maximum RF amplitude at 0\%\ (15~kHz, pulse 4 and 11.26~kHz, Chirp80). Theoretical values for the inversion profile are plotted on the left as a function of resonance offset, with phase deviation $\varphi$ relative to the target $-M_x$ plotted on the right.  Adiabatic Chirp80 produces significant phase errors within the bandwidth for all RF scalings, in contrast to the almost ideal performance of the optimal control \BURBOPinv\ pulse.}
\label{CompareChirp2}
\end{figure*}

\begin{figure*}[h]
\includegraphics[scale=.55]{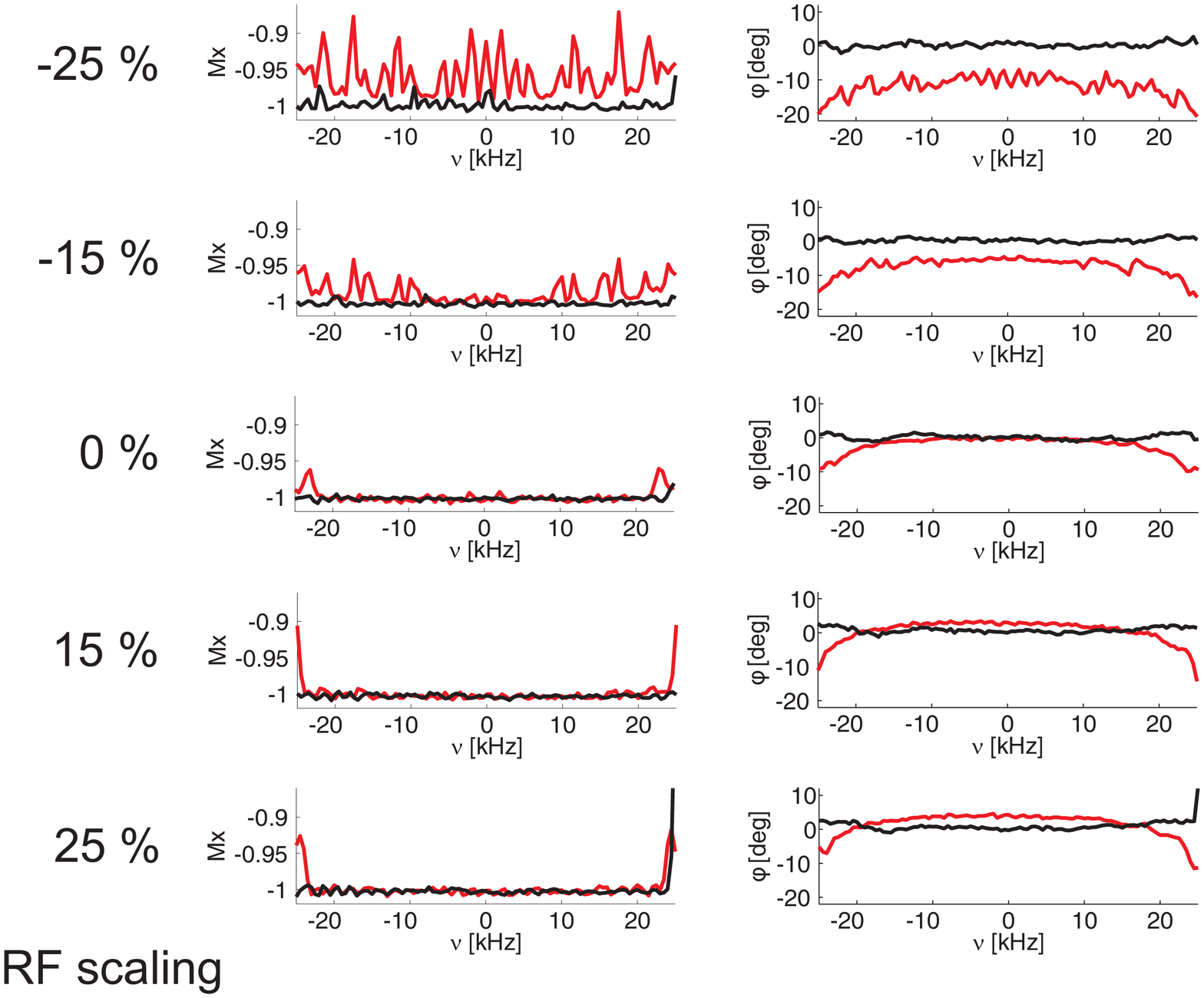}
\caption{Experimental measurements of the inversion profile (left) and phase deviation (right) corresponding to the simulations in \Fig{CompareChirp2}, showing excellent agreement between the experimental and theoretical performance of the pulses.}
\label{CompareChirp3}
\end{figure*}

\begin{figure*}[h]
\includegraphics[scale=.5]{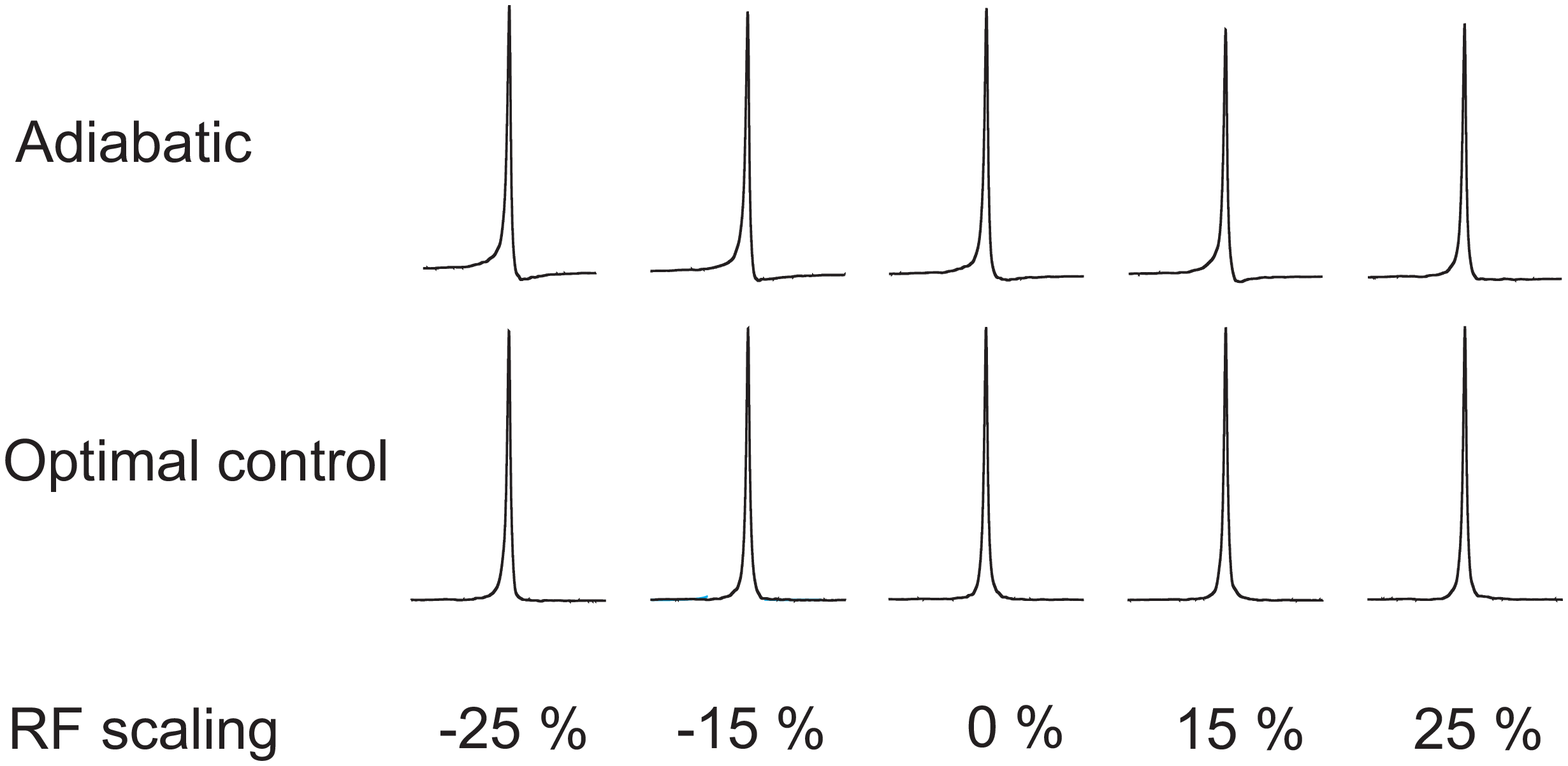}
\caption{Experimental lineshapes comparing optimal control \URinv\ pulse 4 of Table 1 and adiabatic Chirp80 at a resonance offset of $-25$~kHz for RF scalings of $\pm15\%$ and $\pm25\%$ relative to the nominal maximum RF amplitude of 15~kHz at 0\%.  Simulations (not shown) are in excellent agreement with the experiments, as already shown in Figures$\:$\ref{CompareChirp2} and \ref{CompareChirp3}. Chirp80 produces significant experimental phase errors of 20\deg, 15\deg, 9\deg, 11\deg, and 5\deg (reading left-to-right across the RF scalings in the figure) in contrast to the experimental performance of 0.7\deg, 0.7\deg, 0.9\deg, 2.6\deg, 2.5\deg\ for the optimal control \BURBOPinv\ pulse. }
\label{CompareChirp4}
\end{figure*}

\Figure{CompareChirp} compares theoretical performance of pulses 1 and 4 from Table 1 to the Chirp80 pulse. The new pulses significantly improve phase performance over the targeted range of offsets and RF inhomgeneity/miscalibration.  Additional quantitative comparison between pulse 4 and Chirp80 are provided in Figs.\ref{CompareChirp2} and \ref{CompareChirp3}, which also show the excellent agreement between simulations and experimental pulse performance.  Improvements in lineshape and phase that are possible using the new pulses are shown in \Fig{CompareChirp4}. 
\begin{center}
\begin{table*}[ht]
{\small
\hfill{}
\begin{tabular}{c c c c c l l}
\hline
Pulse  & Algorithm & RF$_{\rm nominal}$ & RF$_{\rm max}$ & Inhomogeneity & 
                                       \multicolumn{2}{c}{\underline {Transformation Error}} \\
      &       &  (kHz)          & (kHz)           &   optimization   & (amplitude) & (phase)  \\
\hline
1    & A$_T$     &  11  & 15  & $\pm10\%$  & $< 0.2\%$  & $<2^\circ$   \\
2    & A$_{S,T}$ &  11  & 15  & $\pm15\%$  & $< 0.9\%$  & $<0.3^\circ$   \\
3    & A$_{S,T}$ &  10  & 15  & $\pm15\%$  & $< 1.5\%$  & $<0.5^\circ$  \\
4    & A$_{S,T}$ &  11  & 15  & $\pm25\%$  & $< 2\%$    & $<0.8^\circ$  \\
\hline
\end{tabular}
}
\hfill{}
\caption{Four pulses optimized to execute a 180\deg\ universal rotation about the y-axis over resonance offsets of 50 kHz and RF field inhomogeneity listed in column 5. All pulses are 2~ms long.  For all but 60\musec, pulse RF is constant at the value RF$_{\rm nominal}$, increasing to RF$_{\rm max}$ in the middle of the pulse (see \Fig{Pulse4}).  Amplitude errors in the transformation are the maximum deviation from the target magnetization over the optimized ranges of resonance offset and RF inhomogeneity, expressed as a percent.  Similarly, phase errors represent the maximum deviation from the target phase.  Performance of pulses 1 and 4 are shown in more detail in \Fig{CompareChirp}}
\label{tb:tablename}
\end{table*}
\end{center}


\section{Experiment}

All experiments were implemented on a
Bruker 750~MHz Avance III spectrometer equipped with SGU units
for RF control and linearized amplifiers, utilizing a triple-resonance PATXI probehead and gradients along the $z$-axis.  Measurements are the residual HDO signal in a sample of 99.96\% D$_2$O doped with CuSO$_4$ to a \T{1} relaxation time of 100~ms at 298\deg\ K. 
Signals were obtained at offsets between $-25$~kHz to 25~kHz in steps of 500~Hz. To demonstrate the tolerance of the pulses to RF inhomogeneity/miscalibration, the experiments were repeated with RF amplitude incremented by $\pm 15\%$ and $\pm 25\%$ relative to the nominal maximum RF amplitude for each pulse (15~kHz for pulse 4 of Table 1, 11.26~kHz for Chirp80).  To reduce the effects of RF field inhomogeneity within the coil itself, approximately 40~$\mu$l of the sample solution was
placed in a 5~mm Shigemi limited volume tube.

\section{Conclusion}

We have presented three optimal control algorithms for the de novo design of universal rotation pulses, applied specifically to inversion.  The most noteworthy innovations for NMR spectroscopy are inclusion of the construction principle discovered in \cite{URconstruction} and a time-dependent or ``floating'' limit on the peak RF allowed during the pulse.  The new algorithms result in improved performance compared to existing UR pulses constructed as composites of point-to-point pulses.  
The methodology is very general, and further improvements in the design of robust universal rotation pulses can be anticipated.  
The new pulses are implemented in the Bruker pulse library and will also be made available for downloading from the website http://www.org.chemie.tu-muenchen.de/glaser/Downloads.html.


\section*{Acknowledgments}
T.E.S. acknowledges support from the National Science Foundation under Grant CHE-0943441. B.L. thanks the Fonds der Chemischen Industrie and the Deutsche
Forschungsgemeinschaft (Emmy Noether fellowship LU 835/1-3) for support.  S.J.G. acknowledges support from the DFG (GI 203/6-1), SFB 631 and the Fonds der Chemischen Industrie.  The experiments were performed at the Bavarian NMR Center, Technische Universit\"at  M\"unchen. 

\section{Appendix}
The results of section \ref{FlavorII} were derived as follows.   
\subsection{Euler angles}

The rotation of a vector can be represented as the composition of three elementary (Euler) rotations about the fixed axes used to represent the vector: rotation by angle $\psi$ about the z-axis followed by rotation $\theta$ about the x-axis and rotation $\varphi$ about the z-axis. 
Using the notation $c_\beta = \cos\beta, s_\beta = \sin\beta$, and $R_k(\beta)$ for a rotation by angle $\beta$ about axis $k$, we have
	\begin{eqnarray}
R_z(\psi) & = & 
\left( \begin{array}{ccc}
                       c_\varphi & -s_\varphi & 0 \\
                       s_\varphi & c_\varphi & 0 \\
                       0 & 0 & 1 
               \end{array}  \right ) \nonumber \\
R_x(\theta) & = &
         \left( \begin{array}{ccc}
                      1 & 0 & 0 \\
                      0 & c_\theta & -s_\theta \\
                      0 & s_\theta & c_\theta 
                \end{array}
              \right ) \nonumber \\
R_z(\varphi) & = &
\left( \begin{array}{ccc}
                      1 & 0 & 0 \\
                      0 & c_\theta & -s_\theta \\
                      0 & s_\theta & c_\theta 
       \end{array}
      \right )
	\end{eqnarray}
giving
	\begin{eqnarray}
R & = & R_z(\varphi) R_x(\theta) R_z(\psi) \nonumber \\
   &=&
\left( \begin{array}{ccc}
   c_\psi c_\varphi - c_\theta s_\psi s_\varphi &
  -c_\varphi s_\psi - c_\psi c_\theta s_\varphi &
   s_\theta s_\varphi \\
   c_\theta c_\varphi s_\psi + c_\psi s_\varphi &
   c_\psi c_\theta c_\varphi - s_\psi s_\varphi &
  -c_\varphi s_\theta \\
   s_\psi s_\theta & c_\psi s_\theta & c_\theta  \end{array}
   \right ) \nonumber \\
&=&
\left( \begin{array}{ccc}
   R_{11} & R_{12} & R_{13} \\
   R_{21} & R_{22} & R_{23} \\
   R_{31} & R_{32} & R_{33}  \end{array}
   \right ) 
\label{W90}
	\end{eqnarray}
The Euler angles describing the rotation are thus
	\begin{eqnarray}
\theta &=& \cos^{-1}[\,R_{33}\,] \nonumber \\
\varphi &=& \tan^{-1}[\,R_{13} / (-R_{23})\,] \nonumber \\
\psi &=& \tan^{-1}[\,R_{31} / R_{32}\,]
\label{EulerAngles}
	\end{eqnarray}

Consider the rotation matrix $W$ that produces the \PPexc\ transformation $I_z \rightarrow -I_y$. For initial $I_z$, the first z-rotation has no effect. Angle $\theta$ therefore gives a direct comparison with the desired rotation angle of 90\deg, while $\varphi$ gives the phase deviation from the target state $-I_y$.  $W$ also rotates initial states $I_x$ and $I_y$ to final states that can be arbitrary in the design of the original PP transformation.  This provides a great deal of flexibility in designing PP transformations compared to UR transformations, which must rotate each initial state to a specific final state.  For an ideal \URexc\ rotation about the x-axis, $\psi$ would be zero.  It thus provides the orientations of the rotated vectors $W I_x$ and $W I_y$ in the plane orthogonal to the final state $W I_z$.  

The pulse that generates a \URinv\ transformation from this particular \PPexc\ pulse is constructed from the phase-inverted pulse, denoted by $\overline {90}_{PP}$, followed by the time-reversed (tr) pulse $90^{tr}_{PP}$ \cite{URconstruction}.
At a resonance offset $-\nu$, we have the corresponding rotation operators 
	\begin{eqnarray}
W^{tr}(-\nu) &=& R_z(180)\, W^{-1}(\nu)\: R_z(180) \nonumber \\
            &=&
\left( \begin{array}{ccc}
   W_{11} & W_{21} & -W_{31} \\
   W_{12} & W_{22} &- W_{32} \\
   -W_{13} & -W_{23} & W_{33}  \end{array}
   \right ) 
	\end{eqnarray}
and
	\begin{eqnarray}
\overline{W}(-\nu) &=& R_x(180)\, W(\nu)\: R_x(180) \nonumber \\
            &=&
\left( \begin{array}{ccc}
   W_{11} & -W_{12} & -W_{13} \\
  - W_{21} & W_{22} & W_{23} \\
   -W_{31} & W_{32} & W_{33}  \end{array}
   \right ), 
	\end{eqnarray}
where $R_z(180)$ is zero except for $(-1,-1,1)$ along the diagonal, $R_x(180)$ is similar, but with $(1,-1,-1)$ along the diagonal, and $W^{-1}$ is the transpose of $W$.

The \URinv\ rotation matrix $\widetilde{W}(-\nu) = W^{tr}(-\nu) \overline{W}(-\nu)$ therefore differs from the identity matrix to the extent terms in the product for each matrix element have different signs.  For example, $\widetilde{W}_{11}$ is equal to $W_{11}^2 - W_{21}^2 + W_{31}^2$ rather than $W_{11}^2 + W_{21}^2 + W_{31}^2 = 1$.  Adding and subtracting $W_{21}^2$ in the expression for $\widetilde{W}_{11}$ gives $1 - 2 W_{21}^2$.  Proceding similarly, all the matrix elements can be written in terms of their difference from zero (off-diagonal) or $\pm 1$ (diagonal) to obtain
	\begin{equation}
\widetilde W = \left( \begin{array}{ccc}
   1 - 2W_{21}^2 & 2W_{21} W_{22} & 2 W_{21} W_{23} \\
   -2W_{21}W_{22} & -1 + 2W_{22}^2 & 2W_{23}W_{22} \\
   2W_{21}W_{23} & -2W_{22}W_{23} & -1 + 2(W_{13}^2 + W_{33}^2)  \end{array} \right )
\label{Wtilde}
	\end{equation}
for the \URinv\ rotation operator $\widetilde W$ at resonance offset $-\nu$ in terms of elements $W_{i j}$ of the \PPexc\ rotation operator at offset $\nu$.  Differences in performance at offsets $\pm\nu$ are small to the extent the PP optimization is successful in generating uniform performance over a symmetric range of resonance offsets.

The Euler angles for $\widetilde W$ obtained using \Eq{EulerAngles} give $\tilde\varphi = \tilde\psi$. This result provides geometric insight into the performance of $\widetilde W$.  Rotation about the $z$-axis followed by 180\deg\ rotation about $x$ then rotation by the original amount about $z$ will return $x$ to $x$ and send $y$ and $z$ to $-y$ and $-z$, respectively.  The deviation from the ideal goal of a 180\deg\ rotation determines the errors in the \URinv\ transformation.  For initial $I_x$, the phase angle $\phi$ for the rotated state $\widetilde{W} I_x$ relative to the target state $I_x$ is $\tan\phi = \widetilde{W}_{21}/\widetilde{W}_{11}$.  Using \Eq{W90} and adapting our earlier notation to $\tilde{c}_\beta = \cos\tilde\beta$, etc., gives 
	\begin{eqnarray}
\tan\phi &=& \tilde{c}_\psi \tilde{s}_\psi\,(1 + \tilde{c}_\theta) 
             \over \tilde{c}^2_\psi\, (1 - \tilde{c}_\theta \tan^2 \tilde\psi) \nonumber \\
        &=& \tan\tilde\psi\, {1 + \tilde{c}_\theta
            \over 1 - \tilde{c}_\theta \tan^2 \tilde\psi}
\label{tan(phi)}
	\end{eqnarray}
The extrema for $\tan\phi$ as a function of $\tilde\psi$ occur for $d(\tan\phi)/d\tilde\psi = 0$.  There are two solutions, one of which is an inflection point and the other producing a maximum for $\tan\tilde\psi = -\tilde{c}_\theta^{\,-1/2}$, so that
	\begin{equation}
\tan\phi_\mathrm{max} = {1 + \tilde{c}_\theta \over 2\sqrt{-\tilde{c}_\theta} }
	\end{equation}
According to \Eq{EulerAngles} and \Eq{Wtilde},
	\begin{eqnarray}
\cos\tilde\theta &=& \widetilde{W}_{33} \nonumber \\
 &=& -1 + 2(W_{13}^2 + W_{33}^2) \nonumber \\
 &=& -1 + 2(s_\theta^2\, s_\varphi^2 + c^2_\theta) \nonumber \\
 &=& -1 + 2(c_{\delta\theta}^2\, s_\varphi^2 + s^2_{\delta\theta})
\label{W33tilde} ,
	\end{eqnarray}
where we have substituted $\theta = \pi/2 + \delta\theta$.

Both $\delta\theta$ and $\varphi$ are small for a good \PPexc\ pulse. Expanding $c_{\delta\theta} \approx 1 - (\delta\theta)^2/2$, $s_\varphi \approx \varphi$, $s_{\delta\theta} \approx \delta\theta$, performing the multiplications in \Eq{W33tilde}, and keeping terms to second order gives, with $\tilde\theta = \pi + \delta\tilde\theta$,
	\begin{eqnarray}
\cos\tilde\theta &\approx& -1 + 2\,(\delta\theta^2 + \varphi^2) \nonumber \\
\cos\delta\tilde\theta &\approx& 1 - 2\,(\delta\theta^2 + \varphi^2) \nonumber \\
1 - \delta\tilde\theta^2 / 2 &\approx& 1 - 2\,(\delta\theta^2 + \varphi^2) \nonumber \\
\delta\tilde\theta &\approx& 2 \sqrt{\delta\theta^2 + \varphi^2}.
\label{180RotAngle}
	\end{eqnarray}
Thus, the deviation $\delta\tilde\theta$ from the ideal \URinv\ rotation angle is slightly more than twice the deviation $\delta\theta$ from the ideal \PPexc\ rotation angle.  

Using \Eq{180RotAngle} for $\cos\tilde\theta$, the small-angle approximation for $\tan\phi$, and applying $(1-x)^{1/2} \approx 1 - x/2$ for small $x$ to $\tan\tilde\psi = 1/(-\cos\tilde\theta\,)^{1/2}$ gives
	\begin{eqnarray}
\quad\quad\tan\phi_\mathrm{max} \approx \phi_\mathrm{max} &\lesssim& \delta\theta^2 + \varphi^2 
                                  \over 1 - (\delta\theta^2 + \varphi^2) \nonumber \\
          &\approx& \delta\theta^2 + \varphi^2
\label{PhiMax}
	\end{eqnarray}

Similarly, for initial $I_y$, the phase relative to the target $-y$-axis is $\tan\phi = \widetilde{W}_{12}/\widetilde{W}_{22}$, which results in $\tan\tilde\psi$ being replaced by $\cot\tilde\psi$ in \Eq{tan(phi)}.  The solution for $\tan\phi_\mathrm{max}$ then occurs for  $\cot\tilde\psi = (-\cos\tilde\theta)^{-1/2}$, giving the same bound for $\phi_\mathrm{max}$ as in \Eq{PhiMax}.

\subsection{Angle/axis parameterization}

Alternatively, we can describe a rotation in terms of rotation angle $\eta$ about an axis defined by unit vector $\hat{\bm{n}}$ to obtain
	\begin{eqnarray}
\eta &=& \cos^{-1}{\mathrm{Tr}\, R - 1 \over 2} \nonumber \\
n_x &=& R_{32} -R_{23} \over 2 \sin\eta \nonumber \\
n_y &=& R_{13} -R_{31} \over 2 \sin\eta \nonumber \\
n_z &=& R_{21} -R_{12} \over 2 \sin\eta.
\label{angle-axis}
	\end{eqnarray}
Applying this to $\widetilde{W}$ in \Eq{Wtilde} shows $n_y=0$, as expected from symmetry arguments noted in section \ref{FlavorII}.  The rotation axis $\hat{\bm{n}}$ makes an angle $\alpha$ with respect to the $x$-axis given by $\tan\alpha = n_z/n_x = (\widetilde{W}_{21} - \widetilde{W}_{12}) / (\widetilde{W}_{32} - \widetilde{W}_{23})$.  Following arguments similar to those leading to \Eq{tan(phi)} gives, for small $\alpha$,
	\begin{eqnarray}
\alpha \approx \tan\alpha  &=& 2\tilde{c}_\psi \tilde{s}_\psi\,(1 + \tilde{c}_\theta) 
             \over 2\tilde{c}_\psi \tilde{s}_\theta \nonumber \\
        &\le& (1 + \tilde{c}_\theta)/\tilde{s}_\theta \nonumber \\
        &\approx& \sqrt{\delta\theta^2 + \phi^2}.
\label{tan(alpha)}
	\end{eqnarray}
Here, we have written $\tilde{s}_\theta = (1-\tilde{c}^2_\theta)^{1/2}$, substituted from \Eq{180RotAngle} keeping terms to second order, and used the maximum value of one for $\tilde{s}_\psi$.

We obtain similarly from \Eq{angle-axis} $\cos\eta = -1 + 2W_{22}^2 = \widetilde{W}_{22}$ after rearranging terms as in the discussion preceding \Eq{Wtilde}.  Substituting as above gives $\widetilde{W}_{22} = \tilde{c}^2_\psi \tilde{c}_\theta - \tilde{s}^2_\psi$.  Writing $\tilde{s}^2_\psi = 1 - \tilde{c}^2_\psi$ and employing the maximum value of one for $\tilde{c}_\psi$ gives
	\begin{equation}
\cos\eta \le \cos\tilde\theta 
	\end{equation}
and the deviation of each angle from the ideal rotation of 180\deg\ can be relatively quantified as
	\begin{equation}
\delta\eta \le \delta\tilde\theta,
	\end{equation}
with $\delta\tilde\theta$ given in terms of the \PPexc\ angles in \Eq{180RotAngle}.

\vfill\eject

\end{document}